\documentclass{ws-procs9x6}
\def\bea{\begin{eqnarray}}
\def\eea{\nonumber\end{eqnarray}}
\def\be{\begin{equation}}
\def\ee{\nonumber\end{equation}}

\begin{document}
\title{ANALYTIC CALCULATION OF \\ 
	BOSE-EINSTEIN CORRELATIONS AT LEP1 AND LEP2
\footnote{\uppercase{D}edicated to the memory of 
\uppercase{B}. \uppercase{A}ndersson.
\uppercase{T}his work was partially
supported by an \uppercase{NWO-OTKA} grant \uppercase{N}25487,
by \uppercase{OTKA} grants \uppercase{T038406} and 
\uppercase{T026435}, by a \uppercase{NATO} Senior Science Fellowship
and by the \uppercase{MTA-OTKA-NSF} grant \uppercase{INT0098462}. }}
\author{T. CS\"ORG\H{O}}
\address{MTA KFKI RMKI,
H-1525 Budapest 114, P.O.Box 49, Hungary\\ 
E-mail: csorgo@sunserv.kfki.hu}

\maketitle
\abstracts{
Theoretical results in the $\tau$-model are discussed
in the context of Bose-Einstein correlations in $e^+e^-$ annihilations
at LEP1 and LEP2 as well as that of relativistic heavy ion collisions at
SPS and RHIC. 
}

A current motivation\cite{lonnblad}
to study of Bose-Einstein\cite{andersson} (or HBT\cite{HBT1}
or GGLP\cite{GGLP}) correlations
in $e^+ e^-$ annihilations at LEP1 or LEP2 is to 
determine their possible effects on the mass reconstruction
of $W$ bosons at LEP2, see e.g. ref.\cite{kittel-rev} for a recent review.
The goal is to progress in 2 steps\cite{kittel-rev}:

1) understand Bose-Einstein correlations  at LEP1, at the  $Z^0$ peak, 

2) generalize these results to the case of  $W^+W^-$ production at LEP2.

\section{Bose-Einstein correlations in $\tau$-model at LEP1 }
A model of strongly correlated phase-space 
was developed in ref\cite{qinv-cstjz}  
to explain the experimentally found  invariant relative momentum  
$Q=\sqrt{-(k_1 - k_2)^2}$ dependence of Bose-Einstein
correlations (BEC-s) in $e^+e^-$ reactions\cite{tasso1,tasso2}.
As the correlation function was expressed 
in terms of a distribution of a proper-time $\tau$,
let us call this model the $\tau$-model\cite{qinv-cstjz,be-lep1,be-lep2}.
It is based on 

\underline{ Assumption {\it i}):}
The momentum $k$ and the average position $\overline{x}$ 
are strongly correlated :
$
	\overline{x} = d\, k 
$
where $ d$ is a constant of proportionality, specified below,
and $x \equiv x^\mu = (t,{\bf r})$, $k \equiv k^\mu = (E_{\bf k},{\bf k})$.

\underline{Assumption {\it ii}):} 
The space-time momentum-space correlation is narrower 
than the proper-time distribution 
(when both are measured in dimensionless units).

	Thus the emission function of the $\tau$-model is
\be
	S(x,k) = \int_0^\infty H(\tau) \delta_\Delta(x - d\, k)  N_1(k)
\ee
	where $H(\tau)$ is the proper-time distribution,
	the factor $\delta_\Delta(x - d\, k)$ is  
	describing the strength of the correlations between coordinate
	space and momentum space variables. In the original version
	of the $\tau$-model, these correlations were taken to be ideally
	strong,    $\delta_\Delta(x - d\, k) 
	= \delta^{(4)}(x - d\, k) $, a four-dimensional Dirac-delta.
	The results of the $\tau$-model are unchanged\cite{be-lep1}
	if the $(x,k)$ correlations are described by a sufficiently, 
	but not infinitely narrow function, 
	whose width is characterized by some dimensionless and 
	 narrow scale $\Delta \ll \Delta\tau/\overline{\tau}$,
	where  $\Delta\tau$ is the width of $H(\tau)$. 
	The distributions are normalized,
$
	\int_0^\infty H(\tau)  = 1, 
$
and 
$
	\int d^4 x \delta_\Delta(x - \tau k/m)  = 1.
$

If the expansion is 1+1 dimensional, then $d = \tau_l/m_t$,
if the expansion is 1+3 dimensional then  $d = \tau/m$,  where
the (longitudinal) proper-time is defined as 
($\tau_l =  \sqrt{t^2 - r_z^2}$ and) 
$\tau =  \sqrt{t^2 - r_x^2 - r_y^2 - r_z^2}$, while
the (transverse) mass is given as
($m_t=\sqrt{E_{\bf k}^2 - {\bf k}_z^2}$  and)
$m=\sqrt{E_{\bf k}^2 - {\bf k}^2}$. 
The former is relevant in 2-jet decays
of $Z^0$ at LEP1, the latter is relevant in the case
of fully hadronic $W^+W^-$ decays, mostly 4-jet events.

The  experimentally measurable single-particle spectra  
	$N_1(k)$ is 
\be
	\int d^4 x S(x,k) = N_1(k) 
\ee
arbitrary, and can be taken directly from the measurements.

	The two-particle BEC-s 
	$C_2(k_1,k_2)  =  {N_2(k_1,k_2)}/({N_1(k_1) N_1(k_2)})$
	are calculated with the help of the 
	Yano-Koonin formula\cite{qinv-cstjz,be-lep1,be-lep2}.
	The key step is that for any choice of $d$, one gets 
$
	(\overline{x}_1 - \overline{x}_2)	(k_1 - k_2)
		 = - 0.5(d_1 +d_2) Q^2.
$
	Using a saddle-point integration
	based on {\it i)} and {\it ii)},
	we find that
\be
	C_2(k_1,k_2)  \simeq  
	1 + \lambda {\mathit{Re}}[\tilde H^2(w)],
\ee
	where the argument is   $w = Q^2/(2 m)$ for 1+3 dimensional,
	and $w = Q^2/(2 m_t)$ for 1+1 dimensional expansions,
	corresponding to the appropriate choice for variable $d$,
	and 
$
	\tilde H(w) = \int_0^\infty d\tau H(\tau) \exp(i w\tau) 
$
	is a Fourier-transform of $ H(\tau)$.
	Thus an invariant relative momentum $Q$ dependent BEC appears, 
	if the $(x,k)$ correlations between spacetime and momentum-space 
	are strong enough and the proper-time
	distribution is broad enough.

	For example, let  us consider two-jet events at LEP1. 
	These are 1+1 dimensionally
	expanding systems, hence $d = \tau_l/m_t$. 
	In case of the  asymmetric L\'evy $H(\tau)$
	distribution, $H(\tau) \propto \frac{1}{(\tau-\tau_0)^{3/2}}
\exp(- \frac{4 \Delta\tau}{\tau-\tau_0} )$, we get\cite{qinv-cstjz}
\be
C(Q ) = 1 + \cos(R_0^2 Q^2) \exp(-Q R) \label{e:levy}
\ee
with $R_0^2 = \tau_0/m_t$ and $R^2 = \Delta\tau/m_t$.
This form  is similar to the observed data at LEP1.
Currently known\cite{kittel-rev} 
experimental constraints on BEC-s in $e^+e^-$ annihilations at LEP1 are 
summarized below.  It is easy to show that the $\tau$-model satisfies 
all these properties\cite{be-lep1,be-lep2}: 
\begin{enumerate}
\item{} BEC-s exist in both $Z^0$  decays at LEP1
	and in fully hadronic decays of $W^+W^-$ events at LEP2. 
\item{} The pion source is not spherically symmetric, but elongated
	along the thrust axis, when analyzed in the LCMS frame.
\item{} The shape of the BEC
	function is far from Gaussian at LEP1. 
\item{} The effective source size decreases as $1/\sqrt{m}$ for pions, kaons,
	protons and $\Lambda$ particles,
        and  as $1/\sqrt{m_t}$ for pions with different $m_t$. 
\item{} The normalized 
	three-particle cumulant correlation coincides with 1
	within experimental errors in a large region of $Q_3$.
\item{} No {\it significant} isospin dependence is seen.
\item{} In $e^+ + e^-$ annihilations, the two-particle 
	BEC depends on the relative momentum components only through
	the invariant momentum difference 
	$Q = \sqrt{- (k_1 - k_2)^2} $.
\item{} After the peak, the BEC develops a shallow  
	$C(Q) < 1 $ dip region.
\end{enumerate}

\section{Bose-Einstein correlations from the $\tau$-model at LEP2 }
	The key step is to introduce $\tau$ for $e^+e^- \rightarrow
	W^+W^- $ fully hadronic decays. This variable is 
	now measured from the production point of the $W^+W^-$ pair
	(and not from the decay point of any of the $W$-bosons).

	In these reactions the particle emitting source has two
	components, 
\be
	S(x,k) = S^+(x,k) + S^-(x,k),
\ee
	corresponding to the decays of $W^+$ and $W^-$, respectively.
	Thus the single particle  spectrum has also two components,
$
	N_1(k) = N_1^+(k) + N_1^-(k).
$
	When calculating the BEC-s for such
	a binary source, the key observation is 
$
	H^+(\tau) = H^-(\tau) \equiv H_2(\tau) .
$
	This simple equation has dramatic consequences 
	for the BEC 
 	of fully hadronic decays of $W^+W^-$ pairs at LEP2: 
\be
	C_2(k_1,k_2)  =  
	1 + \lambda {\mathit{Re}}[\tilde H^2_2(\frac{Q^2 }{2m}) ],
\ee
	where 
$
	\tilde H_2(w) = \int_0^\infty d\tau H_2(\tau) 
	\exp(i w\tau) \equiv \tilde H^{\pm}(w) 
$
	is the Fourier-transformed proper-time distribution at LEP2.
	If $H_2(\tau) \approx H_1(\tau)$, where the latter stands for
	the proper-time distribution of particle production 
	in $e^+e^-\rightarrow Z^0$ hadronic decays at LEP1, then 
	the BEC-s at LEP1 and
	LEP2 are also approximately the same.
	This result implies {\it no genuine inter-$W$ Bose-Einstein
	correlations} within the $\tau$-model,
	but the existence of a fully chaotic source with full symmetrization,
	see ref\cite{be-lep2} for further discussions.

\section{Implications for heavy ion collisions}
	The BEC-s or HBT correlations at SPS and at RHIC were mostly
	analysed in terms of the Bertsch-Pratt (BP) or side-out-long variables,
	see refs\cite{cs-rev,Csorgo:2001jf} for
	recent reviews.  
	If we assume that the $\tau$-model and strong $(x,k)$ correlations 
	are relevant also in heavy ion collisions, 
	we can decompose the resulting correlation functions 
	in the BP variables to find a surprizing result:
	The variable of the correlation function,
	$Q^2 \Delta\tau/m_t = Q_s^2 R_s^2 + Q_o^2 R_o^2 + Q_l^2 R_l^2$
	yields
	$R_s^2 = \Delta\tau/m_t$, $R_l^2 = \Delta\tau/m_t$ and 
	$R_o^2 = m^2 \Delta\tau/m_t^3$, where subscripts 
	$\{\null_s, \null_o, \null_l\}$ 
	stand for side, out and long, respectively.

	Experimentally, this implies
	$R_s \approx R_l \propto 1/\sqrt{m_t}$, 
	and $R_o$ decreases with $m_t$ faster than these. 
	Thus very strong $(x,k)$ correlations in the $\tau$-model
	generate $R_o < R_s$ dynamically!
	More detailed studies are needed to determine the proper
	variables of BEC-s in high energy heavy ion collisions.
	If the $\tau$-model is relevant in high energy heavy ion collisions, 		it may imply that $R_o \ll R_s$ means a very {\it broad} 
	proper-time distribution and strong $(x,k)$ correlations.
	Experimentally, this question can be decided  by determining if
	the Bose-Einstein correlation function at RHIC depends 
	on the relative momentum components only through the invariant 
	relative momentum $Q$, as in the case
	of $e^+ e^-$ collisions, or not.

\vfill\eject\end{document}